\documentclass[reprint,superscriptaddress,amsmath,amssymb,amsfonts,aps,prb,floatfix,twocolumn]{revtex4-2}
\usepackage{times}
\usepackage{graphicx}
\usepackage{dcolumn}
\usepackage{bm}
\usepackage{ulem}
\usepackage[colorlinks=true, citecolor=blue, linkcolor=blue, urlcolor=blue]{hyperref}
\usepackage{bbold}
\DeclareMathAlphabet\mathbfcal{OMS}{cmsy}{b}{n}
\usepackage{wasysym}
\usepackage{amssymb}
\usepackage{amsmath}
\usepackage[usenames,dvipsnames]{xcolor}
\usepackage{tikz}

\begin{document}

\title{Entropy-Enhanced Fractional Quantum Anomalous Hall Effect}
\author{Gal Shavit}
\affiliation{Department of Physics and Institute for Quantum Information and Matter, California Institute of Technology,
Pasadena, California 91125, USA}
\affiliation{Walter Burke Institute of Theoretical Physics, California Institute of Technology, Pasadena, California 91125, USA}

\begin{abstract}
Strongly interacting electrons in a topologically non trivial band may form exotic phases of matter.
An especially intriguing example of which is the fractional quantum anomalous Hall phase, recently discovered in twisted transition metal dichalcogenides and in moir\'e graphene multilayers.
However, it has been shown to be destabilized in certain filling factors at sub-100 mK temperatures in pentalayer graphene, in favor of a novel integer quantum anomalous Hall phase [Z. Lu \textit{et al.}, \href{https://arxiv.org/abs/2408.10203}{arXiv:2408.10203}].
We propose that the culprit stabilizing the fractional phase at higher temperatures is its rich edge state structure.
Possessing a multiplicity of chiral modes on its edge, the fractional phase has lower free energy at higher temperatures due to the excess edge modes entropy.
We make distinct predictions under this scenario, including the system-size dependency of the fractional phase entropic enhancement, and how the phase boundaries change as a function of temperature.
\end{abstract}

\maketitle

\textit{Introduction.}
The fractional quantum anomalous Hall effect~\cite{FCIbernevigPRX,FCIneuportPRL} (FQAH) is the zero magnetic field analog of the celebrated fractional quantum Hall effect~\cite{IntroFQHexp2,IntroFQLaughlin}.
Recent advances in moir\'e materials have recently enabled experimental observation of this topological strongly-correlated phase of matter in twisted transition metal dichalcogenides~\cite{FQAHseattle,FCIcornell,FCItransportSeattle,FCItransport_Shanghai}, and crystalline graphene/hBn moir\'e superlattices~\cite{pentalayer_lu2023fractional,FCIantiFCI_xie2024evenodddenominatorfractionalquantum,Young_FQAH_choi2024electricfieldcontrolsuperconductivity}.

A recent experiment~\cite{LongJu_experiment_lu2024extendedquantumanomaloushall} revealed an unexpected surprise: throughout much of the phase diagram in Ref.~\cite{pentalayer_lu2023fractional}, the observed FQAH phases at various filling factors are \textit{not} the actual ground state of the system.
When the electronic temperature was lowered, a different phase superseded the FQAH.
This phase is characterized by vanishing longitudinal resistivity and quantized anomalous Hall response $R_{xy}\approx h/e^2$.
Moreover, it extends to a finite range of densities, with little regards to commensuration with the underlying moir\'e structure.
It has thus been dubbed the extended integer quantum anomalous Hall phase (EIQAH).

To understand the root cause of this phenomenon, a better understanding of both the FQAH phases and the EIQAH is required.
A recent proposal~\cite{SenthillFQAHtransitions_patri2024extendedquantumanomaloushall} has associated the EIQAH with a Wigner-crystal like phase formed on top of an integer quantum anomalous Hall state.
An attractive feature of this theory is that it accounts for some of the non-linear transport signatures in Ref.~\cite{LongJu_experiment_lu2024extendedquantumanomaloushall}.
However, the temperature-driven transition into the FQAH remains unresolved.
Ref.~\cite{sarma2024thermalcrossovercherninsulator} posited that the ``leftover'' electrons in the EIQAH Anderson-localize due to disorder.
Heating the EIQAH delocalizes these electrons, therefore allowing interaction effects to take root, such that the system crosses over to a FQAH phase.

\begin{figure}
    \centering
    \includegraphics[width=9cm]{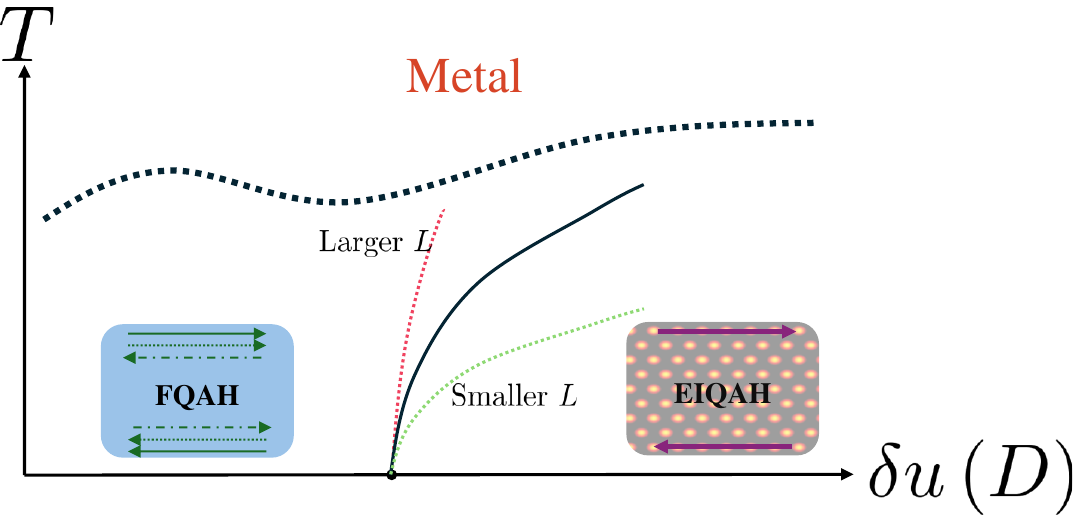}
    \caption{
    Schematic description of the phase diagram emerging from our model.
    The ground state of the system is tuned by some parameter, e.g., displacement field $D$, determining the energy difference between the fractional quantum anomalous Hall phase (FQAH) on the left side, and the extended integer quantum anomalous Hall phase (EIQAH) on the right.
    The left and right insets illustrate a possible edge picture of the respective phases.
    The integer EIQAH phase a single chiral edge mode, whereas the fractional phase may host a rich plethora of edge modes, each contributing to the overall entropy.
    At higher temperatures, the phase boundary (dark blue solid line) shifts as $T\propto\sqrt{\delta u}$  [see Eq.~\eqref{eq:transitioncondition}], favoring the phase with higher edge-state entropy.
    As the linear dimension of the system $L$ increases, the shift of the boundary becomes more subtle (red dotted line) and the entropy driven transition may be preempted by a transition to a  metallic phase.
    }
    \label{fig:schematic}
\end{figure}

In this Letter, we propose a different theoretical explanation for the observed phenomenon, relying only on the topological distinction between the two competing phases, and on the mesoscopic character of the experimental devices.
Namely, excess gapless edge modes in the FQAH phase endow it with a higher entropy.
This naturally results in a phase boundary which favors the FQAH over the EIQAH as temperature increases, see Fig.~\ref{fig:schematic}.

It is rather suggestive that the temperature-stabilized FQAHs appear at filling factors where the analogous fractional quantum Hall phases have rich non-universal edge content~\cite{Macdonald_edge_reconstruct,KFP_23_Reconstruction_PhysRevLett.72.4129,Meir_Edge_reconstruct,Gefen_meir_edge_reconstruct}.
Whereas one naturally expects the EIQAH to host one chiral edge state, at a FQAH filling of, e.g., $\sim 2/3$, one may encounter as many as four downstream/upstream modes~\cite{KaneFisherThermalHallPhysRevB.55.15832,Heiblum_reconstructionSabo2017} which carry heat and thus contribute to the overall edge entropy~\cite{pendry1983quantumEntropy}.

We provide key predictions which may confirm or invalidate the proposed scenario.
Namely, under our assumptions the entropic effect is necessarily device-size-dependent, and should be virtually undetectable for a large enough sample.
Moreover, the temperature dependence of the phase boundary between the phases should follow $T\propto\sqrt{\delta u}$, with $\delta u $ the ground state energy difference between the phases.
Finally, the appreciable evolution of the EIQAH-FQAH boundary as a function of electric displacement field and temperature suggests that this $\delta u$ should be relatively insensitive to displacement field changes.

\textit{Theory.}
We aim to provide a phenomenological characterization of the competition between the extended quantum anomalous Hall phase ($I$) and the fractional quantum anomalous Hall phase ($F$), via a description of the free energy associated with these phases.
Assuming a temperature scale far below the respective transition temperatures of these phases to a metallic phase, we approximate the free energies as,
\begin{equation}
    G_{\mu}\left(x,T\right) = U_\mu\left(x\right)-TS^{\rm edge}_{\mu},\label{eq:F_mu}
\end{equation}
with the index $\mu=I,F$ for the appropriate phase.
Here, $x$ is a parameter tuning a phase transition between the integer and fractional phases.
In the experiment, this parameter is, e.g., an electric displacement field.

Let us denote by $\Omega$ and $\mathtt P$ the area and perimeter of the device.
We will define the potential energy density difference between the two phases
$\delta u \equiv \left(U_F-U_I\right)/\Omega$, which is a phenomenological parameter of our theory.
It is determined by details of the underlying microscopic theory of both the fractional and integer phases, which are beyond the scope of this work.
The entropy density of the phase $\mu$ originating in its linearly dispersing edge states is
\begin{equation}
    s^{\rm edge}_\mu =S^{\rm edge}_\mu/{\mathtt P}=\frac{\pi^2k_B^2T}{3\hbar}
    \sum_{i\in\mu}\frac{1}{v_i},\label{eq:edgeentropy}
\end{equation}
where $v_i$ are the edge mode velocities in the phase $\mu$, and $k_B$ is the Boltzmann constant.
It is then useful to define the velocity scale
\begin{equation}
    v^* = \left[
    \sum_{i\in F}\frac{1}{v_i}-
    \sum_{j\in I}\frac{1}{v_j}\right]^{-1},\label{eq:velocity_star}
\end{equation}
where we assume $v^*>0$, implying the fractional phase edge states carry excess entropy as compared to the integer phase.
If all velocities are equal to $\bar v$, $v^*=\bar{v}/N_\delta$, with $N_{\delta}$ the difference in the number of edge states between $F$ and $I$.
Notice that $v^*$ is dominated by the lowest velocities in the system (and hence by modes carrying the highest density of states).
Furthermore, in the presence of multiple edge modes (as hypothesized for the quantum Hall analogs for the relevant filling factors in the experiment), inter-mode interactions tend to strongly renormalize the velocities~\cite{velocity_renormalized_by_interactions_Braggio_2012,velocity_renormalized_10.21468/SciPostPhys.3.2.014}.
Generically, one or several of the renormalized velocities will be significantly lower than the bare ones.

We note it has been well established that non-linearities in the edge spectrum arise for FQH phases,
at energy scales which are a fraction of the FQH gap~\cite{Jain_nonlinear1,Jain_nonlinear2}.
These non-linearities soften the edge mode dispersion, give rise to an enhancement of the edge state density of states, and consequently lead to a higher entropy than the simplified description in Eq.~\eqref{eq:edgeentropy}.
Assuming similar physics persist for the FQAH edge states, this will make the described entropy-driven effect even more pronounced.

The transition between the two phases, which is of first-order by construction (we consider the competition between the $G_\mu$ described in Eq.~\eqref{eq:F_mu}, is then given by the condition~\cite{SM},
\begin{equation}
    k_B \Tilde{T} =
    \sqrt{\alpha_{\rm geo.} L \hbar v^*}
    \times\sqrt{\delta u\left(\Tilde{x}\right)},\label{eq:transitioncondition}
\end{equation}
where we have conveniently defined the geometrical constant
$3\Omega/\left(\pi^2{\mathtt P}\right)\equiv\alpha_{\rm geo.}L$, relating the surface-to-perimeter ratio of the device to its linear extent $L$ (for, e.g., a square, $\alpha_{\rm geo.}=3/\left(4\pi^2\right)$).

Taking into account realistic parameters from the experiment in Ref.~\cite{LongJu_experiment_lu2024extendedquantumanomaloushall}, one may arrive at a rough estimate of $\delta u$ given Eq.~\eqref{eq:transitioncondition} characterizes the transition.
Considering an electronic density of $n_e\approx 0.5$ 10$^{12}$ cm$^{-2}$, and parameters $v^*\sim10^3$ m/sec, $\Tilde{T}\sim0.1$ K, $L\sim 1$ $\mu$m, the difference in energies between the fractional and extended integer phase $\delta u/\left(k_B n_e\right)$ is on the order of $O$(5 mk / electron). 

Let us take a different perspective, considering specifically the out-of-plane electric displacement field as controlling the transition, $x=D$.
We may formulate the evolution of the phase boundary in the $D$-$T$ plane as a Clausius-Clapeyron relation,
\begin{equation}
    \frac{dD}{dT}=\frac{S^{\rm edge}_F-S^{\rm edge}_I}{\delta \mu_{D}},\label{eq:Clausius}
\end{equation}
where $\delta \mu_{D}/\Omega = \left(\partial\delta u\right)/\left(\partial D\right)$
is the difference of out-of-plane electric dipole moment between the fractional and integer phases.
Eq.~\eqref{eq:Clausius} then implies that the effect we describe here, stabilization of the phase with richer edge content by the edge entropy, is greatly enhanced \textit{if the energy difference between the phases is insensitive to changes in the displacement field}.
As the relevant experimental regime lies in the high-field region with $D\approx1$ V/nm, this is actually quite likely -- the displacement field effects on the single-particle properties of the relevant band are nearly saturated.

\begin{figure}
    \centering
    \includegraphics[width=8.7cm]{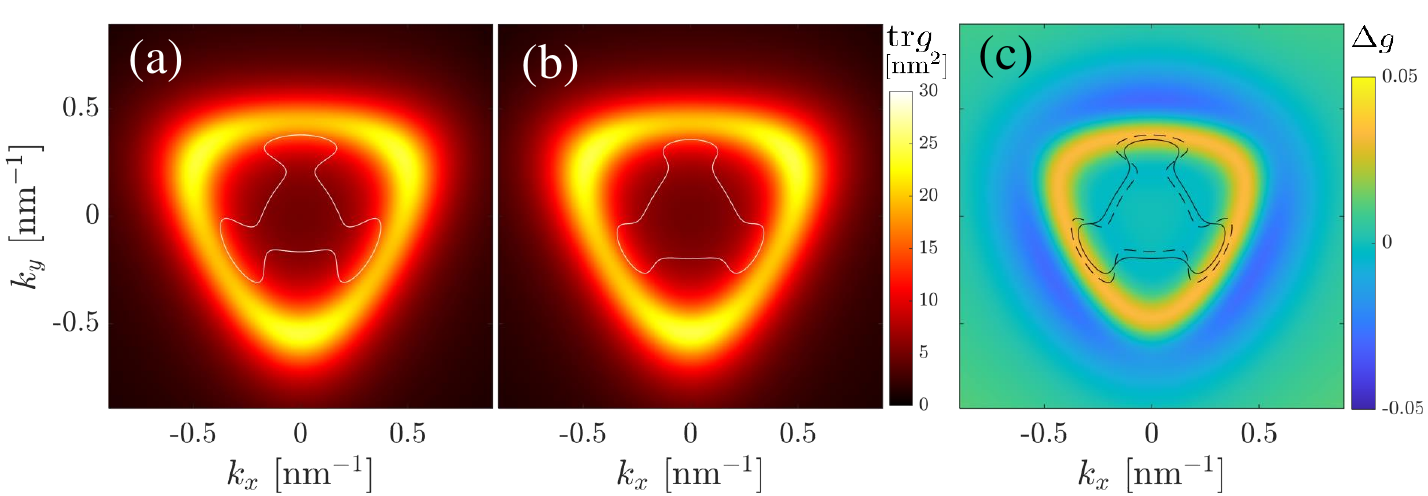}
    \caption{
    Insensitivity of the quantum geometry to displacement field.
    (a)
    Trace of the Fubini-Study metric around the valley $K$ point for an interlayer potential difference of 40 meV.
    The white contour marks the Fermi surface corresponding to the density $n_e=6$ $\cdot$10$^{11}$ cm$^{-2}$.
    (b)
    Same as (a), with interlayer potential difference of 38 meV.
    (c)
    The relative difference between (a) and (b), which is significantly less than the relative potential change.
    Dashed and solid lines correspond to the Fermi surfaces in (a) and (b), respectively.
    Details of the calculations for this figure appear in the Supplementary Materials~\cite{SM}.
    }
    \label{fig:pentalyertraceg}
\end{figure}

An in-depth study of the effect of $D$ on the stability of the various FQAH states, and on its energetic competition with the EIQAH phase (whose nature is yet unresolved) is beyond the scope of this work.
Such a study requires of course, among other things, consideration of the electron-electron interactions necessary to facilitating a fractional phase.
Let us, however, examine a single-particle property which has been shown to play a key role in stabilizing the fractional quantum anomalous Hall phase in a given band, namely the trace of the Fubini-Study metric ${\rm tr} g$~\cite{RoytraceconditionFCI,BergholtzLatticeConstantBerry,FCI_Khalef_moire,FCI_TBG_parker2021fieldtuned,simonrudnerindicators,Jackson2015Roynumerical,BergholtzLatticeConstantBerry,FCI_TBG_parker2021fieldtuned,numerical_tmd_pressure_PhysRevResearch.5.L032022,TMD_numerical_Devakul,TMD_numerical_wang2023fractional,shavit2024FCI}. 
For simplicity, we omit the effect of the moir\'e-inducing hBN layer, since electrons are mostly polarized to the graphene layer far away from it in the relevant regime.
A change of 5\% to the potential difference between the outermost graphene layers hardly has any effect on this quantum geometric property, as seen by comparing Fig.~\ref{fig:pentalyertraceg}a--b.
In Fig.~\ref{fig:pentalyertraceg}c we plot the relative change of this metric, which remains much smaller than the relative change in $D$ throughout the majority of the enclosed Fermi volume at the relevant densities.
This simple calculation demonstrates that at the experimentally relevant values of displacement field, a decisive parameter controlling the FQAH physics remains quite insensitive to changes in $D$.
Such insensitivity is this consistent with a very small $\delta\mu_D$.


In the thermodynamic limit, $L\to\infty$, and the entropy-driven transition temperature diverges.
Referring again to Fig.~\ref{fig:schematic}, the phase boundary between the FQAH and EIQAH phases becomes nearly vertical below the transition temperature to some gapless phase.
Practically, one expects to ``lose'' the transition once a device exceeds a certain size $\Tilde{L}$, for which the transition temperature exceeds the temperature at which the fractional phase forms, $\Tilde{T}>T_c$.
This length scale is approximately
\begin{equation}
    \Tilde{L}\approx
    \frac{\left(k_BT_c\right)^2}{\alpha_{\rm geo.} \hbar v^* \delta u}.\label{eq:lengthscale}
\end{equation}
Using the estimate for $\delta u$ we obtained above, given $T_c$ which is of order of a few hundred mK, one may estimate $\Tilde{L}\sim O\left(10\,\mu{\rm m}\right)$.
If the excess edge density of states ($\propto1/v^*$) is large enough, or the potential energy difference $\delta u$ is rather small, the EIQAH-to-FQAH ``melting'' can be observed for realistic device sizes.

We note that in the presence of disorder, specifically of the kind that couples differently to the $F$ and $I$ phases, the impact of the edge-state entropy may be further \textit{enhanced}.
Domain walls separating the topologically distinct phases contain protected edge modes which carry entropy at finite temperature.
A transition from a uniform $I$ phase at low-temperatures to a domain-patterned phase at higher temperature, analogous to the so-called Chern mosaic~\cite{Chern_mosaic_Grover2022}, is expected in such a scenario.
In such a mosaic phase, $\alpha_{\rm geo.}$ effectively becomes much smaller, since one replaces the perimeter of the device $\tt P$, with the overall length of domain walls in the system, roughly proportional to $\tt P$ times the number of domains.
The interplay between disorder and domain wall entropy driven transitions has been thoroughly explored in Ref.~\cite{Shavit_domains_PhysRevLett.128.156801}.
The signature this sort of transition would have on \textit{transport} is not universal, would depend on the position of the domains relative to the contacts, and would thus be hard to reproduce. 

\textit{Discussion.}
We have presented a scenario accounting for the emergence of FQAH upon heating of an EIQAH ground state.
The culprit in this scenario is the excess entropy of the FQAH edge states.
If the ground-state energy difference between the phases is sufficiently small, the system may gain energy by ``partially melting'' in its edges.
This is enabled by transitioning into the intermediate FQAH, which allows additional gapless excitations on its edge.

Clearly, the free energy gain on the edge is expected to be eclipsed by the bulk $\delta u>0$ contribution for large enough systems, and thus the behavior of the phase boundary will be strongly system-size-dependent.
This is the first distinct prediction of our model, indicating that \textit{the movement of the phase boundary in the $D$--$T$ plane should be inversely proportional to $L$}, as inferred by Eq.~\eqref{eq:Clausius}.

As the transition is dominated by edge contributions, our theory further predicts this effect should be enhanced (i.e., the FQAH more stable against the EIQAH) at filling fractions which host a larger abundance of edge states.
We carefully speculate that this effect may explain why certain FQAHs disappear completely at low temperatures, whereas others survive for a finite range of displacement fields.
An exhaustive mapping of the hierarchy of the stabilized FQAHs at intermediate temperatures is necessary in order to make this point more concrete.

In addition to experimental predictions, our proposal raises issues to be addressed by the theory underlying the FQAH-EIQAH competition.
Namely, in the picture described above the energetic difference between the phase $\delta u$ is rather small.
Moreover, $\delta u$ should be relatively insensitive to changes of the displacement field around the experimentally relevant range, $D\approx$ 1 V/nm.
These points may provide a valuable clue towards disentangling the two phases, as well as a testable hypothesis in their numerical simulations.

The predictive power of Eq.~\eqref{eq:transitioncondition}, suggesting the phase boundary takes the shape $\Tilde{T}\propto\sqrt{\delta u\left(\Tilde{x}\right)}$, is somewhat diminished by the lack of knowledge of the functional dependence of $\delta u\left({x}\right)$.
Improved theoretical understanding of the phase competition will help resolve this issue and provide an ideal setting to test Eq.~\eqref{eq:transitioncondition}.
Alternative explanations resulting in the same square-root behavior are possible,
as the nature of collective excitations in FQAH and EIQAH is far from understood.
These include the anomalous magneto-roton~\cite{GMP_magnetoroton_PhysRevB.33.2481,Repellin_magnetoroton_PhysRevB.90.045114}, which is gapped, and presumably contributes very little entropy at low temperatures.
Phonon excitations due to spontaneous crystallization in the EIQAH are expected to be gapped due to the weak moir\'e potential~\cite{SenthillFQAHtransitions_patri2024extendedquantumanomaloushall}, and even if they are not (or the gap is very small), the phonons should give an entropic advantage to the EIQAH and not the FQAH.
We note that the observed FQAH itself may harbor an additional hidden translation-symmetry-breaking order~\cite{FQAHScrystallu2024fractionalquantumanomaloushall}, yet the resulting phonons are also expected to be gapped for the same reason -- the underlying moir\'e lattice.


In conclusion, the rich landscape of FQAH, EIQAH, anomalous Hall crystals, and their interplay in moir\'e flat-band systems is far from being understood.
The anomalous appearance of the exotic FQAH phase at higher temperatures is one of many observations calling to question our understanding of these materials.
Drawing inspiration from the presumably-similar physics of the fractional quantum Hall effect, we illustrate how this phenomenon can be related to the \textit{topological difference} between the FQAH and the EIQAH ground states, regardless of their intricate collective excitation spectra.

\hfill

G.S. acknowledges enlightening discussions with Gil Refael, as well as support from the Walter Burke Institute for Theoretical Physics at Caltech, and from the Yad Hanadiv Foundation through the Rothschild fellowship.

\bibliography{FQAHentropy}

\end{document}